%% file: lat2021suzuki_hiraguchiMOD_1025.tex
\documentclass[a4paper,11pt]{article}
\usepackage{amsmath,latexsym}
\usepackage{amssymb}
\usepackage{pos}
\usepackage{multirow}
\input revmacro
\def\Tr{\text{Tr}} 
\newcommand{\F}{\phantom {1}}
\def\nn{\nonumber}
\newcommand{\beqn}{\begin{eqnarray}}
\newcommand{\beq}{\begin{equation}}
\newcommand{\eeqn}{\end{eqnarray}}
\newcommand{\eeq}{\end{equation}}

\newcommand{\kbra}[1] { \left< #1 \right>}

\title{Monopoles of the Dirac type and color confinement in QCD}

\author*[a]{Tsuneo Suzuki}
\author*[b,c]{Atsuki Hiraguchi}
\note{T.S.:Speaker of Sections 1, 2, 3 and 5, A.H.: Speaker of Section 4}
\author[d]{Katsuya Ishiguro}
\affiliation[a]{RCNP, Osaka University,\\
Ibaraki, Osaka 567-0047, Japan}
\affiliation[b]{Institute of Physics, National Yang Ming Chiao Tung University,\\
Hsinchu 30010, Taiwan}
\affiliation[c]{Department of Mathematics and Physics, Kochi University, \\ 
Kochi 780-8520, Japan}
\affiliation[d]{Library and Information Technology, Kochi University,\\
Kochi 780-8520, Japan}

\emailAdd{tsuneo@rcnp.osaka-u.ac.jp}
\emailAdd{a.hiraguchi@nycu.edu.tw} 
\emailAdd{ishiguro@kochi-u.ac.jp}

\abstract{We present results of $SU(3)$ Monte-Carlo studies of a new color confinement scheme proposed recently due to Abelian-like monopoles of the Dirac type corresponding  in the continuum limit to violation of the non-Abelian Bianchi identities (VNABI). The simulations are done without any additional gauge-fixing smoothing the vacuum. We get for the first time, in pure $SU(3)$ simulations with the standard Wilson action, (1) the perfect Abelian dominance with respect to the static potentials on $12^4\sim 16^4$ lattices at $\beta=5.6-5.8$ using the multilevel method. (2) The perfect monopole as well as Abelian dominances with respect to the static potentials by evaluating the Polyakov loop correlators on $24^3\times4$ at $\beta=5.6$. The Abelian photon part gives zero string tension. (3) The Abelian dual Meissner effect is observed with respect to the Abelian gauge field and Abelian monopoles. The Abelian electric field of a color is squeezed due to the solenoidal monopole current with the corresponding color. Although the scaling and the volume dependence are not yet studied in $SU(3)$, the present results and the previous $SU(2)$ results are consistent with the new Abelian picture of color confinement that each one of eight (three in $SU(2)$) colored electric flux  is squeezed by the corresponding colored Abelian-like monopole of the Dirac type corresponding to VNABI.
}

\FullConference{
 The 38th International Symposium on Lattice Field Theory, LATTICE2021
  26th-30th July, 2021
  Zoom/Gather@Massachusetts Institute of Technology
}

\begin{document}
\maketitle
\section{Introduction}
Color confinement in  quantum chromodynamics (QCD) 
is still an important unsolved  problem~\cite{CMI:2000mp}. 

As a picture of color confinement, 't~Hooft~\cite{tHooft:1975pu} and Mandelstam~\cite{Mandelstam:1974pi} conjectured that the QCD vacuum is a kind of a magnetic superconducting state caused by condensation of magnetic monopoles and  an effect dual to the Meissner effect works to confine color charges. This conjecture is very interesting, but there are many problems to be unsolved even at the present stage. In contrast to SUSY QCD~\cite{Seiberg:1994rs} or Georgi-Glashow model~\cite{'tHooft:1974qc,Polyakov:1976fu} with scalar fields, to find color magnetic monopoles  is not straightforward in QCD. 
 
Without scalar fields, it seems necessary to introduce some singularities  
 as shown by Dirac~\cite{Dirac:1931} in $U(1)$ quantum electrodynamics. 
 An interesting idea to introduce such a singularity is to project QCD to the Abelian 
maximal torus group by a partial (but singular) gauge fixing~\cite{tHooft:1981ht}. In $SU(3)$ QCD, the maximal torus group is  Abelian $U(1)^2$. Then color magnetic monopoles appear as a topological object at the space-time points corresponding to the singulariry of the gauge-fixing matrix. Condensation of the monopoles  causes  the dual Meissner effect with respect to $U(1)^2$.
Numerically, an Abelian projection in various gauges such as the maximally Abelian (MA)
gauge~\cite{Kronfeld:1987ri,Kronfeld:1987vd} seems to support the conjecture~\cite{Suzuki:1992rw, Chernodub:1997ay}.
  
Although numerically interesting, the idea of Abelian projection~\cite{tHooft:1981ht} is theoretically very unsatisfactory. 1) In non-perturabative QCD, 
any gauge-fixing is  not necessary at all. There are infinite ways of such a partial gauge-fixing and whether the 't Hooft scheme is gauge independent or 
not is not known. 2)  Especially, consider a Polyakov gauge in which  Polyakov loops are diagonalized. In this gauge, the Abelian dual Meissner picture works good~\cite{Sekido:2007mp}, where space-like monopole currents play the role of the solenoidal current squeezing the electric field. However, this fact contradicts the 'tHooft idea that monopoles appear at the space-time points where the eigenvalues become degenerate. In the Polyakov loop gauge, such a degenerate point runs only in the time-like direction.  Hence monopoles predicted in the 'tHooft idea should always be time-like. (3)   After an Abelian projection, only one (in $SU(2)$) or two (in $SU(3)$) gluons are photon-like with respect to the residual $U(1)$ or $U(1)^2$ symmetry and the other gluons are massive charged matter fields. Such an asymmetry among gluons is  unnatural. Also it is not clear enough that the 'tHooft picture is sufficient for non-Abelian color (not Abelian charge) confinement.

In 2010  Bonati et al.~\cite{Bonati:2010tz} found an interesting fact that the violation of non-Abelian Bianchi identity (VNABI) exists behind the Abelian projection scenario in various gauges and hence gauge independence is naturally expected. This is completely different from the original 'tHooft idea of monopoles. Along this line,   one of the authors (T.S.)~\cite{Suzuki:2014wya} found a more general relation that VNABI is just equal to the violation of Abelian-like Bianchi identities corresponding to the existence of Abelian-like monopoles. A partial gauge-fixing is not necessary at all from the beginning. If the non-Abelian Bianchi identity is broken, Abelian-like monopoles necessarily appear due to a line-like singularity leading to a non-commutability with respect to successive partial derivatives. This is hence an extension of the Dirac idea of monopoles in QED to non-Abelian QCD. 

\vspace{.5cm}
In this report, (1) the new theoretical scheme for color confinement based on the dual Meissner effect due to the above monopoles is summarized shortly. (2) The first results showing the perfect Abelian dominance and the monopole dominance in pure $SU(3)$ lattice QCD are shown next along with the short review of pure $SU(2)$ results~\cite{Suzuki:2007jp,Suzuki:2009xy}. (3) The results showing the existence of the  dual Abelian Higgs mechanism in pure $SU(3)$  are discussed. (4) Finally numerical results showing the continuum limit of the new monopoles~\cite{Suzuki:2017lco,Suzuki:2017zdh} are reviewed shortly in the framework of pure $SU(2)$ lattice QCD, since existence of the continuum limit is essentially important for the new confinement scheme.

\section{Equivalence of VNABI and Abelian-like monopoles}
First of all, we prove that the Jacobi identities of covariant derivatives lead us to conclusion that  violation of the non-Abelian Bianchi identities (VNABI) $J_{\mu}$ is nothing but an Abelian-like monopole $k_{\mu}$ defined by violation of the Abelian-like Bianchi identities without gauge-fixing. 
 Define a covariant derivative operator $D_{\mu}=\partial_{\mu}-igA_{\mu}$. The Jacobi identities are expressed as 
\begin{eqnarray}
\epsilon_{\mu\nu\rho\sigma}[D_{\nu},[D_{\rho},D_{\sigma}]]=0. \label{eq-Jacobi}
\end{eqnarray}
By direct calculations, one gets
\begin{eqnarray*}
[D_{\rho},D_{\sigma}]&=&[\partial_{\rho}-igA_{\rho},\partial_{\sigma}-igA_{\sigma}]\\
&=&-ig(\partial_{\rho}A_{\sigma}-\partial_{\sigma}A_{\rho}-ig[A_{\rho},A_{\sigma}])+[\partial_{\rho},\partial_{\sigma}]\\
&=&-igG_{\rho\sigma}+[\partial_{\rho},\partial_{\sigma}],
\end{eqnarray*}
where the second commutator term of the partial derivative operators can not be discarded, since gauge fields may contain a line singularity. Actually, it is the origin of the violation of the non-Abelian Bianchi identities (VNABI) as shown in the following. The non-Abelian Bianchi identities and the Abelian-like Bianchi identities are, respectively: $D_{\nu}G^{*}_{\mu\nu}=0$ and $\partial_{\nu}f^{*}_{\mu\nu}=0$.
The relation $[D_{\nu},G_{\rho\sigma}]=D_{\nu}G_{\rho\sigma}$ and the Jacobi identities (\ref{eq-Jacobi}) lead us to
\begin{eqnarray}
D_{\nu}G^{*}_{\mu\nu}&=&\frac{1}{2}\epsilon_{\mu\nu\rho\sigma}D_{\nu}G_{\rho\sigma} \nn\\
&=&-\frac{i}{2g}\epsilon_{\mu\nu\rho\sigma}[D_{\nu},[\partial_{\rho},\partial_{\sigma}]]\nn\\
&=&\frac{1}{2}\epsilon_{\mu\nu\rho\sigma}[\partial_{\rho},\partial_{\sigma}]A_{\nu}\nn\\
&=&\partial_{\nu}f^{*}_{\mu\nu}, \label{eq-JK}
\end{eqnarray}
where $f_{\mu\nu}$ is defined as $f_{\mu\nu}=\partial_{\mu}A_{\nu}-\partial_{\nu}A_{\mu}=(\partial_{\mu}A^a_{\nu}-\partial_{\nu}A^a_{\mu})\lambda^a/2$. Namely Eq.(\ref{eq-JK}) shows that the violation of the non-Abelian Bianchi identities, if exists,  is equivalent to that of the Abelian-like Bianchi identities.

Denote the violation of the non-Abelian Bianchi identities as  $J_{\mu}$:
\begin{eqnarray}
J_{\mu} = \frac{1}{2}J_{\mu}^a\sigma^a
=D_{\nu}G^*_{\mu \nu}. \label{nabi}
\end{eqnarray}
 An Abelian-like monopole $k_{\mu}$ without any gauge-fixing is defined as the violation of the Abelian-like Bianchi identities:
\begin{eqnarray}
k_{\mu}=\frac{1}{2}k_{\mu}^a\sigma^a = \partial_{\nu}f^*_{\mu\nu}
=\frac{1}{2}\epsilon_{\mu\nu\rho\sigma}\partial_{\nu}f_{\rho\sigma}. \label{ab-mon}
\end{eqnarray}
Eq.(\ref{eq-JK}) shows that
\begin{eqnarray}
J_{\mu}=k_{\mu}. \label{JK}
\end{eqnarray}

Due to the antisymmetric property of the Abelian-like field strength, we get Abelian-like conservation conditions~\cite{Arafune:1974uy}:
\begin{eqnarray}
\partial_\mu k_\mu=0. \label{A-cons}
\end{eqnarray}

A few comments are in order.
\begin{enumerate}
\item Eq.(\ref{JK}) can be considered as 
a special case 
of the important relation derived by Bonati et al.~\cite{Bonati:2010tz} in the framework of an Abelian projection to a simple case without any Abelian projection.
\item The Abelian-like conservation relation (\ref{A-cons}) gives us eight conserved magnetic charges in the case of color $SU(3)$ and $N^2-1$ charges in the case of color SU(N). But these are kinematical  relations coming from the derivative with respect to the divergence of an antisymmetric tensor~\cite{Arafune:1974uy}.  The number of conserved charges is different from that of the Abelian projection scenario~\cite{tHooft:1981ht}, where only $N-1$ conserved charges exist in the case of color SU(N).
\end{enumerate}

\section{Abelian static potentials  in $SU(3)$}
In $SU(2)$ QCD, perfect Abelian dominance is proved  without performing any addtional gauge fixing using the multilevel method~\cite{Luscher:2001} in Ref.~\cite{Suzuki:2007jp, Suzuki:2009xy}. Also perfect monopole dominance is proved very beautifully by applying the random gauge transformation as a method of the noise reduction  of measuring gauge-variant quantities in the same reference~\cite{Suzuki:2007jp, Suzuki:2009xy}. However,  in $SU(3)$ QCD on lattice,  it is not straightforward from the beginning. First, to extract Abelian link fields for all eight colors separately from non-Abelian gauge field matrix is not simple, since in $SU(3)$ the non-Abelian gauge field is not expanded by the Lie-Algebra elements in a simple way as in $SU(2)$. We choose the following method to define the Abelian link field by maximizing the following overlap quantity
\begin{eqnarray}
R^a=\sum_{s,\mu}\Tr\left(e^{i\theta_\mu^a(s)\lambda^a}U_\mu^{\dag}(s)\right), \label{RA}
\end{eqnarray}
where $\lambda^a$ is the Gell-Mann matrix and 
sum over $a$ is not taken.
This choice in $SU(2)$ leads us to the same Abelian link fields adopted in  Ref.~\cite{Suzuki:2007jp, Suzuki:2009xy}.

For example, we get from the maximization condition of  (\ref{RA}) an Abelian link field $\theta_1(s,\mu)$ corresponding to $\sigma_1$ ($SU(2)$) and $\lambda_1$ ($SU(3)$) as 
\begin{eqnarray*}
\theta_1(s,\mu)&=&\textrm{tan}^{-1}\frac{U_1(s,\mu)}{U_0(s,\mu)},\ \ \ \ (\textrm{SU2}: \ \ U(s,\mu)=U_0(s,\mu)+i\vec{\sigma}\cdot\vec{U}(s,\mu))\\
&=& 
\textrm{tan}^{-1}\frac{Im(U_{12}(s,\mu)+U_{21}(s,\mu))}{Re(U_{11}(s,\mu)+U_{22}(s,\mu))}, \ \ (\textrm{SU3})
\end{eqnarray*}

Once Abelian link variables are fixed, we can extract Abelian, monopole and photon parts from  the Abelian plaquette variable as follows:
\begin{eqnarray}
\theta_{\mu\nu}^a(s) &=&\bar{\theta}_{\mu\nu}^a(s)+2\pi
n_{\mu\nu}^a(s)\ \ (|\bar{\theta}_{\mu\nu}^a|<\pi),\label{abel+proj}
\end{eqnarray}
where $n_{\mu\nu}^a(s)$ is an integer
corresponding to the number of the Dirac string.
Then an Abelian monopole current is defined by
\begin{eqnarray}
k_{\mu}^a(s)&=& -(1/2)\epsilon_{\mu\alpha\beta\gamma}\partial_{\alpha}
\bar{\theta}_{\beta\gamma}^a(s+\hat\mu) \nonumber\\
&=&(1/2)\epsilon_{\mu\alpha\beta\gamma}\partial_{\alpha}
n_{\beta\gamma}^a(s+\hat\mu)  \label{eq:amon}.
\end{eqnarray}
The current (\ref{eq:amon})  satisfies the Abelian conservation condition (\ref{A-cons}) and takes an integer value which corresponds to the magnetic charge obeying the Dirac quantization condition~\cite{DeGrand:1980eq}.
 
\begin{figure}[htbp]
\centering
\includegraphics[width=7cm]{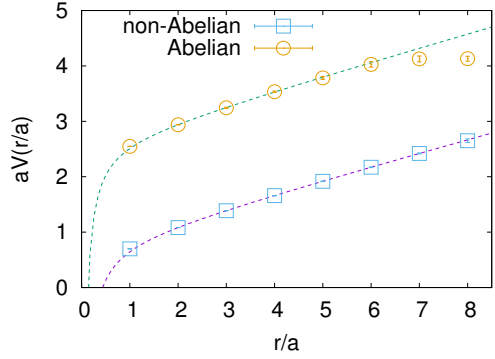}
\vspace*{.5cm}
\caption{The static-quark potentials from non-Abelian and Abelian Polyakov loop correlators at $\beta = 5.60$ on $16^3\times 16$ lattice.}\label{POTENTIAL_16xx4_b560}
\end{figure}
  
\subsection{Perfect Abelian dominance in $SU(3)$}
Now let us evlaluate Abelian static potentials through Polyakov loop correlators written by the Abelian link variable defined above:
\begin{eqnarray}
P_{\rm A} = \exp[i\sum_{k=0}^{N_{t}-1}\theta_1(s+k\hat{4},4)]. \label{eq-PA}
\end{eqnarray}
Since the above Abelian Polyakov loop operator without any additional gauge-fixing is defined locally,  the Poyakov loop correlators can be evaluated through the multilevel method~\cite{Luscher:2001}. Contrary to the $SU(2)$ case in Ref.~\cite{Suzuki:2009xy}, we need much more number of internal updates to get meaningful results. The simulation parameters using the standard Wilson action are shown in Table~\ref{Table1}.
\begin{table}[t]
\centering
\caption{\label{SU3multilevel_parameter}
Simulation parameters for the measurement of static potential using multilevel method.
$N_{\rm sub}$ is the sublattice size divided and $N_{\rm iup}$ is the number of internal updates in the multilevel method .}\label{Table1}
\begin{tabular}{c|c|c|c|c|c}
\hline
$\beta$ &$N_{s}^{3}\times N_{t}$& $a(\beta)$~[fm]& $N_{\rm conf}$ & $N_{\rm sub}$ & $N_{\rm iup}$ \\ 
\hline
5.60 & $12^3 \times 12$ & 0.2235\F & 6 & 2 & 5,000,000\\
5.60 & $16^3 \times 16$ & 0.2235\F & 6 & 2 & 10,000,000\\
5.70 & $12^3 \times 12$ & 0.17016\F & 6 & 2 & 5,000,000\\
5.80 & $12^3 \times 12$ & 0.13642\F & 6 & 3 & 5,000,000\\
\hline
\end{tabular}
\end{table}
An example of the static potentials from non-Abelian and Abelian Polyakov loop correlators are shown in Fig.~\ref{POTENTIAL_16xx4_b560}.
\begin{table}[t]
\begin{center}
\caption{
Best fitted values of the string tension $\sigma a^2$, the
Coulombic coefficient $c$, and the constant $\mu a$ for the
potentials $V_{\rm NA}$, $V_{\rm A}$.}
\label{stringtension_multilevel}
\begin{tabular}{l|c|c|c}
\multicolumn{4}{l}{}\\ 
\hline
$\beta =5.6, 12^3\times 12$& $\sigma a^2$ & $c$ & $\mu a$  \\ 
\hline
$V_{\rm NA}$   & 0.2368(1)  & -0.384(1)  & 0.8415(7)   \\ 
$V_{\rm A}$     & 0.21(5)     & -0.6(6)     & 2.7(4)  \\ 
\hline
\multicolumn{4}{l}{$\beta =5.6, 16^3\times 16$} \\ 
\hline
$V_{\rm NA}$  & 0.239(2)  & -0.39(4)  & 0.79(2)  \\ 
$V_{\rm A}$    & 0.25(2)   &  -0.3(1)   & 2.6(1)  \\
\hline
\multicolumn{4}{l}{$\beta =5.7, 12^3\times 12$} \\
\hline
$V_{\rm NA}$  & 0.159(3)  & -0.272(8)  & 0.79(1)  \\ 
$V_{\rm A}$    & 0.145(9)  & -0.32(2)   & 2.64(3) \\ 
\hline
\multicolumn{4}{l}{$\beta =5.8, 12^3\times 12$}\\ 
\hline
$V_{\rm NA}$  & 0.101(3) & -0.28(1)  &  0.82(1) \\ 
$V_{\rm A}$    & 0.102(9)  & -0.27(2)  &  2.60(3) \\
\hline
\end{tabular}
\end{center}
\end{table}
The best fitted values of the non-Abelian and Abelian string tensions are plotted in 
Table~\ref{stringtension_multilevel}.

\begin{figure}[htbp]
  \begin{minipage}[b]{0.5\linewidth}
    \centering
    \includegraphics[keepaspectratio, scale=0.35]{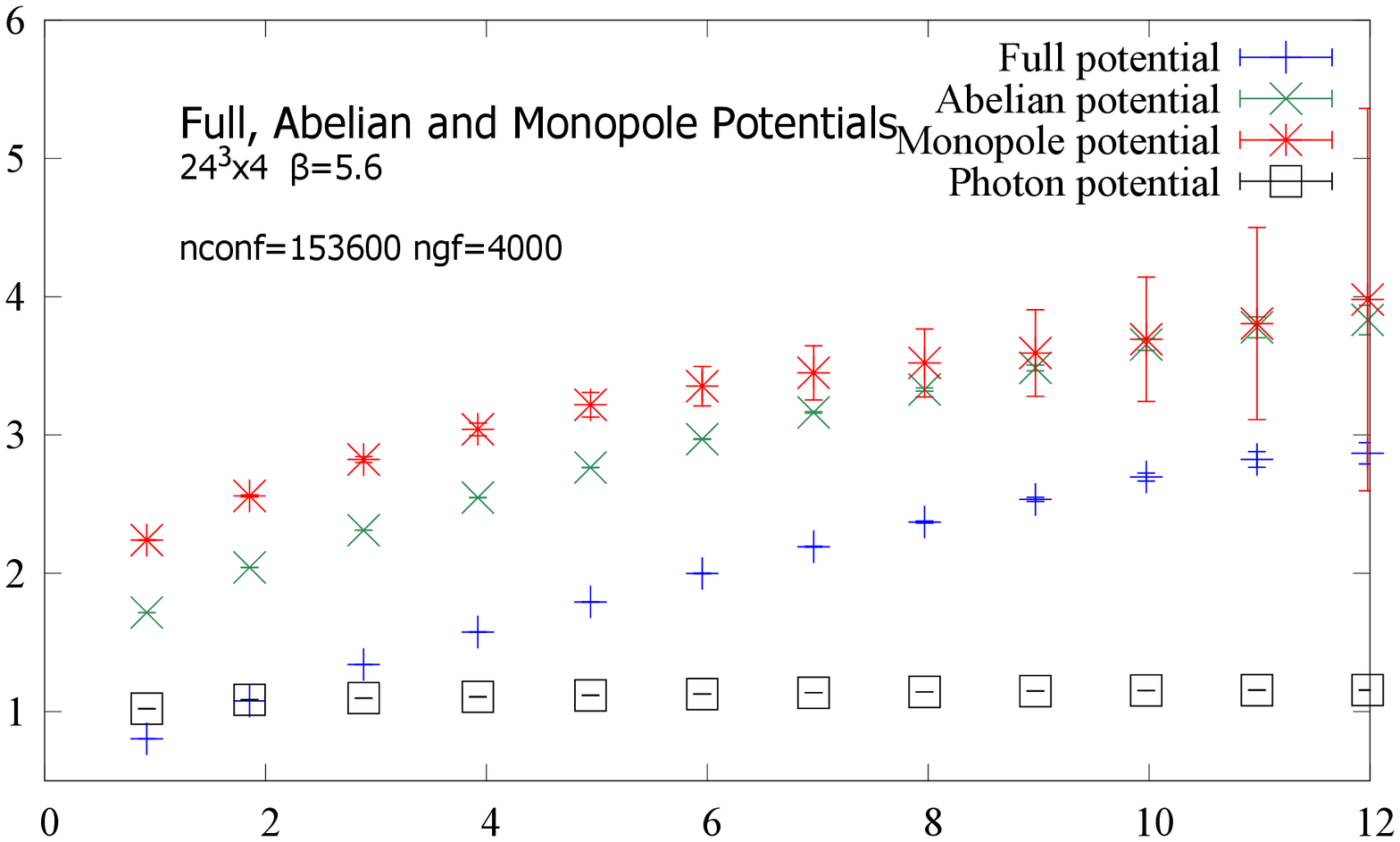}
    \caption{Full, Abelian and Monopole potentials on $24^3\times 4$ at $\beta=5.6$}
  \label{FAMpot_b56}
  \end{minipage}
  \begin{minipage}[b]{0.5\linewidth}
    \centering
    \includegraphics[keepaspectratio, scale=0.35]{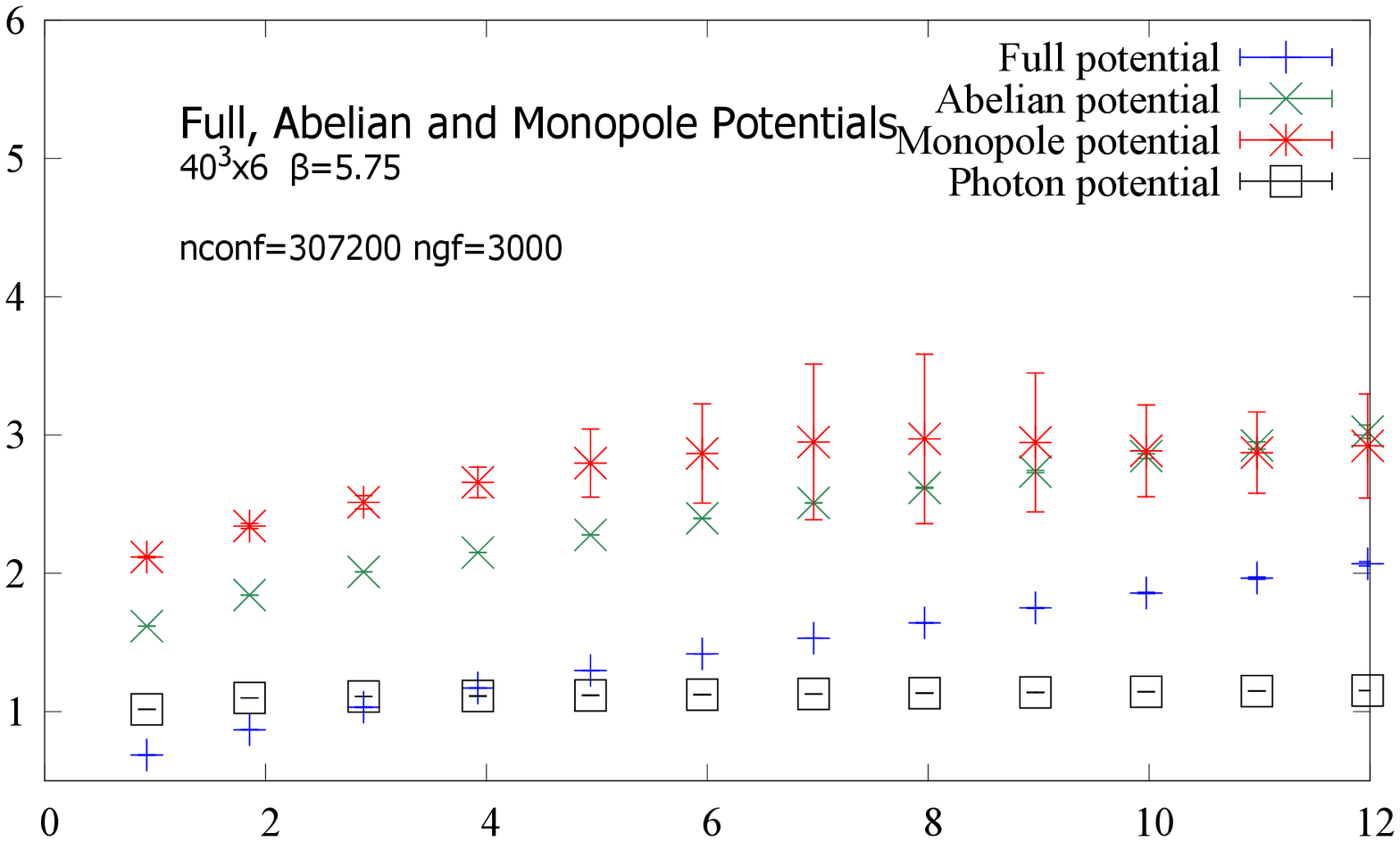}
    \caption{Full, Abelian and Monopole potentials on $40^3\times 6$ at $\beta=5.75$}\label{FAMpot_b575}
  \end{minipage}
\end{figure}

\subsection{Perfect monopole dominance in $SU(3)$}
Without adopting any further gauge fixing smoothing the vacuum, Abelian monopole static potential can reproduce fully the string tension of non-Abelian static potential in $SU(2)$ QCD as shown in Ref.\cite{Suzuki:2007jp, Suzuki:2009xy}. This is called as perfect monopole dominance of the string tension. Almost perfect monopole dominance was found in $SU(3)$ when 
MA gauge is adopted in Ref.\cite{Suganuma:2018}. However, without such additional gauge-fixing, it is found to be tremendously difficult. 

We investigate the monopole contribution to the $SU(3)$ static potential through Polyakov loop correlators in order to examine the role of monopoles for confinement without any additional gauge-fixing.. The monopole part of the Polyakov loop operator is extracted as follows.
Using the lattice Coulomb propagator $D(s-s')$, which satisfies
$\partial_{\nu}\partial'_{\nu}D(s-s') = -\delta_{ss'}$ with a
forward (backward) difference $\partial_{\nu}$ ($\partial'_{\nu}$), 
the temporal component of the Abelian fields $\theta^a_{4}(s)$ are written as 
\begin{equation}
\theta^a_4 (s) 
= -\sum_{s'} D(s-s')[\partial'_{\nu}\theta^a_{\nu 4}(s')+
\partial_4 (\partial'_{\nu}\theta^a_{\nu}(s'))] \; . 
\label{t4}
\end{equation} 
Inserting Eq.~\eqref{t4} to the Abelian Polyakov loop~\eqref{eq-PA},
we obtain
\begin{eqnarray}
&&P^a_{\rm A} = P^a_{\rm ph} \cdot P^a_{\rm mon}\; ,\nonumber\\
&&P^a_{\rm ph} = \exp\{-i\sum_{k=0}^{N_{t}-1} \!\sum_{s'}
D(s+k\hat4-s')\partial'_{\nu}\bar{\theta}^a_{\nu 4}(s')\} \; ,\nonumber\\
&&P^a_{\rm mon} = \exp\{-2\pi i\sum_{k=0}^{N_{t}-1}\! \sum_{s'}
D(s+k\hat4-s')\partial'_{\nu}n^a_{\nu 4}(s')\}\; .\nonumber\\
\label{ph-mon}
\end{eqnarray}
We call $P^a_{\rm ph}$ the photon
and $P^a_{\rm mon}$ the monopole parts of 
the Abelian Polyakov loop $P^a_A$, respectively~\cite{Suzuki:1994ay}.
The latter is due to the fact that the Dirac strings 
$n^a_{\nu 4}(s)$ lead to the monopole currents in 
Eq.~\eqref{eq:amon}~\cite{DeGrand:1980eq}.
Note that the second term of Eq.~\eqref{t4} does
not contribute to the Abelian Polyakov loop 
in Eq.~\eqref{eq-PA}.
We show the simulation parameters and the results in comparison with the $SU(2)$ case.
\begin{table}[t]
\begin{center}
\caption{\label{SU2data}
Simulation parameters for the measurement of the static potential and the force from $P_{\rm A}$, $P_{\rm ph}$ and $P_{\rm mon}$. $N_{\rm RGT}$ is the number of random gauge transformations. }\label{PARA}
\begin{tabular}{c|c|c|c|c}
\hline
$\beta$ &$N_{s}^{3}\times N_{t}$& $a(\beta)$~[fm]& $N_{\rm conf}$ & $N_{\rm RGT}$ \\ 
\hline
$SU2$, 2.20 & $24^{3} \times 4$ & 0.211(7)\F & 6,000 & 1,000\\
2.35 &$24^{3} \times 6$ & 0.137(9)\F & 4,000 & 2,000\\
2.35 &$36^{3} \times 6$ & 0.137(9)\F & 5,000 & 1,000\\
2.43 &$24^{3} \times 8$ & 0.1029(4)& 7,000 & 4,000 \\
\hline
$SU3$, 5.6 & $24^{3} \times 4$ & 0.2235\F & 153,600 & 4,000\\
5.75&$40^3\times 6$& 0.152\F &307,200&3,000\\
\hline
\end{tabular}
\end{center}
\end{table}

\begin{table}
\centering
\fontsize{8pt}{20pt}\selectfont
\caption{
Best fitted values of the string tension $\sigma a^2$, the
Coulombic coefficient $c$, and the constant $\mu a$ for the
potentials $V_{\rm NA}$, $V_{\rm A}$, $V_{\rm mon}$ and $V_{\rm ph}$.
$V_{\rm FA}$ ($V_{\rm FM}$) stands for the potential determined from non-Abelian and Abelian (monopole) Polyakov loop correlators.}\label{FAMPFIT}
\begin{tabular}{l|l|c|c|c|c|c}
 \hline
$SU(2)$& &$\sigma a^2$ & $c$ & $\mu a$ & FR($R/a$) 
& $\chi^2/N_{\rm df}$ \\ \hline
$24^3\times 4$&$V_{\rm NA}$   & 0.181(8)  & 0.25(15) & 0.54(7)  & 3.9 - 8.5 & 1.00 \\ 
&$V_{\rm A}$ & 0.183(8) & 0.20(15) & 0.98(7)  & 3.9 - 8.2 & 1.00 \\ 
&$V_{\rm mon}$ & 0.183(6)  & 0.25(11) & 1.31(5)  & 3.9 - 6.7 & 0.98 \\ 
&$V_{\rm ph}$ &$-2(1)\times 10^{-4}$ & 0.010(1) & 0.48(1)  & 4.9 - 9.4 & 1.02 \\ \hline
$24^3\times 8$&$V_{\rm NA}$  & 0.0415(9)  & 0.47(2) & 0.46(8)  & 4.1 - 7.8\F & 0.99 \\ 
&$V_{\rm A}$ & 0.041(2)  & 0.47(6) & 1.10(3)  & 4.5 - 8.5\F & 1.00 \\ 
&$V_{\rm mon}$ & 0.043(3)  & 0.37(4) & 1.39(2)  & 2.1 - 7.5\F & 0.99 \\ 
&$V_{\rm ph}$ &$-6.0(3)\times 10^{-5}$ &0.0059(3) & 0.46649(6)  & 7.7 - 11.5 & 1.02\\
\hline
\hline
$SU(3)$&&&&&&\\
$24^3\times 4$&$V_{\rm NA}$ & 0.1429(76)  & 0.184(26) & 1.458(90)  & 4 - 12 & 1.18 \\ 
&$V_{\rm A}$ & 0.1840(71)   & 0.482(45)   & 2.749(37)  & 1 - 7 & 1.044 \\
&$V_{\rm FA}$ & 0.1646(66)   & 0.968(149)   & 2.149(63)  & 3 - 11 & 2.17 \\ 
&$V_{\rm FM}$ & 0.172(13)  & 0.303(29) & 2.409(43) & 0.8 - 10 & 0.75 \\ 
$40^3\times 6$&$V_{\rm NA}$   & 0.1034(5)  & 0.411(12) & 0.8700(48)  & 3 - 18 & 0.51 \\ 
&$V_{\rm A}$ & 0.058(31)   & 0.397(72)   & 2.88(10)  & 0 - 6 & 0.38 \\
&$V_{\rm FA}$ & 0.1061(8)   & 0.340(12)   & 1.8216(65)  & 2 - 15 & 0.96 \\
&$V_{\rm FM}$ & 0.1067(10)  & 0.234(23) & 2.275(34) & 0 - 9 & 0.08 \\ 
 \hline
\end{tabular}
\end{table}  
In comparison with those in $SU(2)$ where beautiful Abelian and monopole dominances are observed using reasonable number of vacuum ensembles, we needed much more number of vacuum ensembles even on $24^3\times 4$ small lattice in $SU(3)$ as shown in Tabel~\ref{PARA}. Abelian dominance is seen from Abelian-Abelian Polyakov loop correlators. But in the case of monopole-monopole Polyakov loop correlators, we could not get good results. Since Abelian dominance is seen also from non-Abelian and Abelian Polyakov loop correlators as shown in Tabel~\ref{FAMPFIT}., we try to study non-Abelian and monopole correlators. Since the fit is not good enough, a strong indication of monopole dominance is seen from the hybrid correlators as shown in Fig.~\ref{FAMpot_b56}.
 When we go to larger lattice $40^3\times 6$ at $\beta=5.75$ (correponding to a similar temperature), Abelian dominance is seen using non-Abelian and Abelian correlators. But non-Abelian and monopole correlators are much more worse as shown in Fig.~\ref{FAMpot_b575}. In the case of photon-photon correlators, the string tensions on both cases are almost zero.

\section{The Abelian dual Meissner effect in $SU(3)$}
Let us next discuss the dual Meissner effect due to the Abelian-like monopoles in pure $SU(3)$ QCD.
\subsection{Simulation details of flux-tube profile}
In this section, we show the results with respect to the Abelian dual Meissner effect. In the previous work \cite{Suzuki:2009xy} studying the spatial distribution of color electric fields and monopole currents, they  used the connected correlations between a non-Abelian Wilson loop and Abelian operators in $SU(2)$ gauge theory without gauge fixing. We apply the same method to $SU(3)$ gauge theory without gauge fixing. Here we employ the standard Wilson action on the $24^3(40^3) \times 4$ lattice with the coupling constant $\beta = 5.6$. We consider a finite temperature system at $T = 0.8 T_c$. To improve the signal-to-noise ratio, the APE smearing~\cite{APE:1987} is applied to the spatial links and the hypercubic blocking~\cite{Hasenfratz:2001} is applied to the temporal links.  We introduce  random gauge transformations to improve the signal to noise ratios of the data concerning the Abelian operators. 

\begin{figure}[h]
\begin{center}
\includegraphics[width=4.5cm]{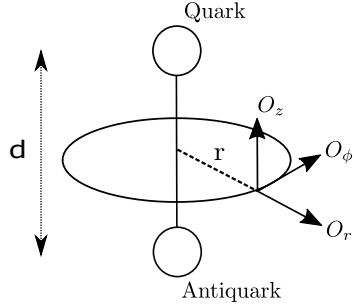}
\end{center}
\caption{The definition of the cylindrical coordinate $(r, \phi, z)$ along the $q\text{-}\bar{q}$ axis. The $d$ corresponds to the distance between Polyakov loops. }\label{cor}
\end{figure}

To measure the flux-tube profiles, we consider the connected correlation functions as done in \cite{Cea2016, refId0}:  
  \begin{align}
  \rho_{conn}(O(r)) = &\frac{\kbra{\Tr(P(0)LO(r)L^{\dagger})\Tr P^{\dagger}(d)}}{\kbra{\Tr P(0) \Tr P^{\dagger}(d)}} 
   - \frac{1}{3}\frac{\kbra{\Tr P(0) \Tr P^{\dagger}(d) \Tr O(r)}}{\kbra{\Tr P(0) \Tr P^{\dagger}(d)}}, \label{connect}
 \end{align} 
where $P$ denotes a non-Abelian Polyakov loop, $L$ indicates Schwinger line, $r$ is a distance from a flux-tube and $d$ is a distance between Polyakov loops. We use the cylindrical coordinate $(r,\phi,z)$ to parametrize the $q\text{-}\bar{q}$ system as shown in Fig \ref{cor}. 

\subsection{The spatial distribution of color electric fields}
\begin{figure}[h]
\begin{center}
\includegraphics[width=7cm]{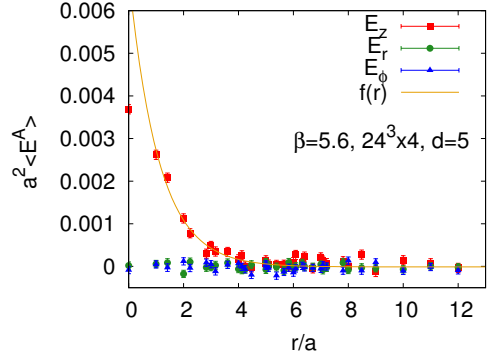}
\end{center}
\captionsetup{skip=24pt}
\caption{The Abelian color electric field with $d=5$ at $\beta = 5.6$ on $24^3\times 4$ lattices.} \label{EA_zrp}
\end{figure}

First of all, we show the results of Abelian color electric fields using an Abelian gauge field $\theta_1(s,\mu)$. To evaluate the Abelian color electric field, we adopt the Abelian plaquette as an operator $O(r)$. We calculate a penetration length $\lambda$ from the Abelian color electric fields with $d=3,4,5,6$ at $\beta=5.6$ and check the $d$ dependence of $\lambda$. To improve the accuracy of the fitting, we evaluate  $O(r)$ at both on-axis and off-axis distances. As a result, we find the Abelian color electric fields $E^A_{z}$ are squeezed as in Fig \ref{EA_zrp}.  We fit these results to a fitting function, 
  \begin{align}
   f(r) = c_1 \mathrm{exp}(-r/\lambda) + c_0 .
  \end{align}  
The parameter $\lambda$ corresponds to the penetration length. We summarize the values of parameters in Table \ref{lambda}. We find the values of the penetration length are almost the same.

\begin{table}[h] 
\begin{center}
\caption{The penetration length $\lambda$ at $\beta = 5.6$ on $24^3 \times 4$ lattices.} \label{lambda}
\begin{tabular}{|c|c|c|c|c|} \hline
 \multicolumn{1}{|c|}{$d$} &  \multicolumn{1}{|c|}{$\lambda/a$}&  \multicolumn{1}{|c|}{$c_1$}  &  \multicolumn{1}{|c|}{$c_0$}&$\chi^2/N_{df}$ \\ \hline
 3&0.91(1)&0.0100(2)&-0.000002(8)&1.31628\\
\hline
4&1.10(6)&0.0077(4)&-0.00005(4)&0.972703 \\
\hline
5&1.09(8)&0.0068(6)&-0.00001(4)&0.995759\\
\hline
6&1.1(1)&0.0055(8)&-0.00008(7)&0.869692 \\
\hline
\end{tabular}
\end{center}
\end{table}


\subsection{The spatial distribution of monopole currents}
Next we show the result of the spatial distribution of Abelian-like monopole currents.  We define the Abelian-like monopole currents on the lattice as in Eq.~(\ref{eq:amon}).
In this study we evaluate the connected correlation (\ref{connect}) between $k^1$ and the two non-Abelian Polayakov loops. We use random gauge transformations to evaluate this correlation. As a result, we find the spatial distribution of monopole currents around the flux-tube at $\beta = 5.6$. Only the monopole current in the azimuthal direction, $k^1_{\phi}$, shows the correlation with the two non-Abelian Polyakov loops.

\begin{figure}[h]
\begin{center}
\includegraphics[width=7cm]{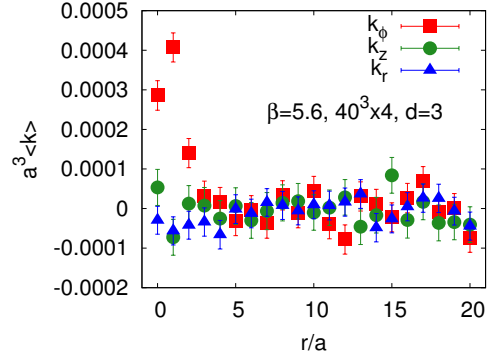}
\end{center}
\captionsetup{skip=24pt}
\caption{The monopole current at $\beta = 5.6$ on $40^3\times 4$ lattices.}
\end{figure}

\subsubsection{The dual Amp\`{e}re's law}
In previous $SU(2)$ researches \cite{Suzuki:2009xy}, they investigated the dual Amp\`{e}re's law to see what squeezes the color-electric field. In the case of $SU(2)$ gauge theory without gauge fixings, they confirmed the dual Amp\`{e}re's law and the monopole currents squeeze the color-electric fields. In this section we show the results of the dual Amp\`{e}re's law in the case of $SU(3)$ gauge theory. The definition of monopole currents leads to the following relation,
\begin{align}
 (\mathrm{rot}E^{a})_\phi = \partial_{t}B^{a}_{\phi} + 2\pi k^{a}_{\phi}, 
\end{align}
where index $a$ is a color index. 

\begin{figure}[h]
\begin{center}
\includegraphics[width=7cm]{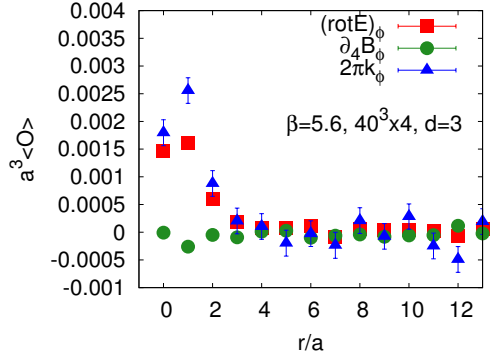}
\end{center}
\captionsetup{skip=24pt}
\caption{The dual Amp\`{e}re's law at $\beta = 5.6$ on $40^3\times 4$ lattices.} \label{dualA}
\end{figure}

As a results, we confirm that there is no signal of the magnetic displacement current $\partial_{t}B^{a}_{\phi}$ aroud the flux-tube with $d=3$ at $\beta=5.6$ as shown in Fig. \ref{dualA}. It suggests that the Abelian-like monopole current squeezes the Abelian color electric field as a solenoidal current in $SU(3)$ gauge theory without gauge fixing, although more data for larger $d$ are necessary.
 
\subsection{The vacuum type in $SU(3)$ gauge theory without gauge fixing}
Finally, we evaluate the Ginzburg-Landau (GL) parameter, which characterizes the type of the (dual) superconducting vacuum. In the previous result \cite{Suzuki:2009xy}, they found that the vacuum type is near the border between the type 1 and type 2 dual superconductors by using the $SU(2)$ gauge theory without gauge fixing. We apply the same method to $SU(3)$ gauge theory. 

\begin{figure}[h]
\begin{center}
\includegraphics[width=7cm]{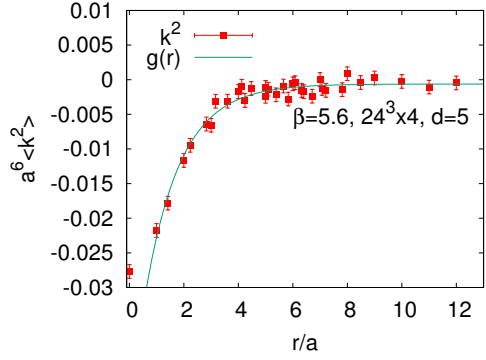}
\end{center}
\captionsetup{skip=24pt}
\caption{The squared monopole density with $d=5$ at $\beta = 5.6$ on $24^3\times 4$ lattices.}\label{kk}
\end{figure}

To evaluate the coherence length, we measure the correlation between the squared monopole density and two non-Abelian Polyakov loops by using the disconnected correlation function~\cite{PhysRevD.72.074505,Suzuki:2009xy},
\begin{align}
 \kbra{k^2(r)}_{q\bar{q}} = &\frac{\kbra{\Tr{P(0)}\Tr{P^{\dagger}(d)}\sum_{\mu, a}k^a_{\mu}(r)k^a_{\mu}(r)}}{\kbra{\Tr{P}(0)\Tr{P^{\dagger}(d)}}} 
 - \kbra{\sum_{\mu,a}k^a_{\mu}(r)k^a_{\mu}(r)}.
\end{align}
We fit the profiles to the function,
\begin{align}
 g(r) = c'_1 \mathrm{exp}\left(-\frac{\sqrt{2}r}{\xi}\right) + c'_0,
\end{align}
where the parameter $\xi$ corresponds to the coherence length. We plot the profiles of $\kbra{k^2(r)}_{q\bar{q}} $ in Fig \ref{kk}.  As a result, we could evaluate the coherence length $\xi$ with $d=3,4,5,6$ at $\beta=5.6$ and find the almost same values of $\xi/\sqrt{2}$ for each $d$. Using these parameters $\lambda$ and $\xi$, we could evaluate the the Ginzburg-Landau parameter. The GL parameter  $\kappa = \lambda/\xi$ can be defined as the ratio of the penetration length and the coherence length. If $\sqrt{2}\kappa < 1$, the vacuum type is of the type 1 and if $\sqrt{2}\kappa > 1$, the vacuum  is of the type 2.  We show the GL parameters in $SU(3)$ gauge theory in Table \ref{GL}. We find that the vacuum  is near the border between type 1 and type 2 or close to the type 1 at one gauge coupling constant $\beta=5.6$. This is the first result of the vacuum type in pure $SU(3)$ gauge theory without gauge fixing, although different $\beta$ data are necessary to show the continuum limit. 
\begin{table}[h] 
\begin{center}
\caption{The coherence length $\xi/\sqrt{2}$ at $\beta = 5.6$ on $24^3 \times 4$ lattices.}
\begin{tabular}{|c|c|c|c|c|} \hline
 \multicolumn{1}{|c|}{$d$} &  \multicolumn{1}{|c|}{$\xi/\sqrt{2}a$}&  \multicolumn{1}{|c|}{$c'_1$}  &  \multicolumn{1}{|c|}{$c'_0$}&$\chi^2/N_{df}$ \\ \hline
3&1.04(6)&-0.050(3)& 0.0001(2)&0.997362 \\
\hline
4&1.17(7)&-0.052(3)&-0.0003(2)&1.01499  \\
\hline
5&1.3(1)&-0.047(3)&-0.0006(3)&0.99758 \\
\hline
6&1.1(1)&-0.052(8)&-0.0013(5)&1.12869 \\
\hline
\end{tabular}
\end{center}
\end{table}
\begin{table}[h] 
\begin{center}
\caption{The Ginzburg-Landau parameters at $\beta=5.6$ on $24^3\times 4$ lattice.}\label{GL}
\begin{tabular}{|c|c|} \hline
 \multicolumn{1}{|c|}{d} &  \multicolumn{1}{|c|}{$\sqrt{2}\kappa$} \\ \hline
3&0.87(5) \\
\hline
4&0.93(7)  \\
\hline
5&0.83(9) \\
\hline
6&0.9(2)  \\
\hline
\end{tabular}
\end{center}
\end{table}

\section{The continuum limit of the new Abelian-like monopoles}
Finally let us review shortly  an important result  showing that the above new-type Abelian monopoles have the continuum limit. The studies in the framework of pure $SU(2)$ QCD were done with respect to the monopole density in  Ref.~\cite{Suzuki:2017lco} and to the effective monopole action in Ref.~\cite{Suzuki:2017zdh}. 

In both studies, it is inevitable to introduce additional gauge fixings to make the lattice vacuum smooth enough reducing the number of lattice artifact monopoles although the gauge dependence problem appear newly. It is due to the fact that even lattice artifact monopoles contribute to the monopole density or the effective monopole action equally. Hence we adopted four different smooth gauge-fixing methods to check gauge dependence, that is, MCG (maximally center gauge)~\cite{DelDebbio:1996mh,DelDebbio:1998uu}, DLCG (direct Laplacian center gauge)~\cite{Faber:2001zs}, MAWL (maximal Abelian Wilson loop gauge)~\cite{Suzuki:1996ax} and MAG+U1 (maximal Abelian gauge~\cite{Kronfeld:1987ri, Kronfeld:1987vd}
 and $U(1)$ Landau gauge).  To make the vacuum smooth, we also introduce a tadpole improved action and the block-spin transformation of monopoles~\cite{Ivanenko:1991wt, Shiba:1994}.
 
Original monopoles are defined on a $a^3$ cube and the $n$-blocked monopoles are defined on a cube with a lattice spacing $b=na$ as follows:
\begin{eqnarray*}
k_{\mu}^{(n)}(s_n) = \sum_{i,j,l=0}^{n-1}k_{\mu}(ns_n+(n-1)\hat{\mu}+i\hat{\nu}
     +j\hat{\rho}+l\hat{\sigma}).
\end{eqnarray*}
We considered $n=1,2,3,4,6,8,12$ blockings  for $\beta=3.0\sim 3.9$ on $48^4$ lattice.

\begin{figure}[h]
    \centering
    \includegraphics[width=8cm,height=6.cm]{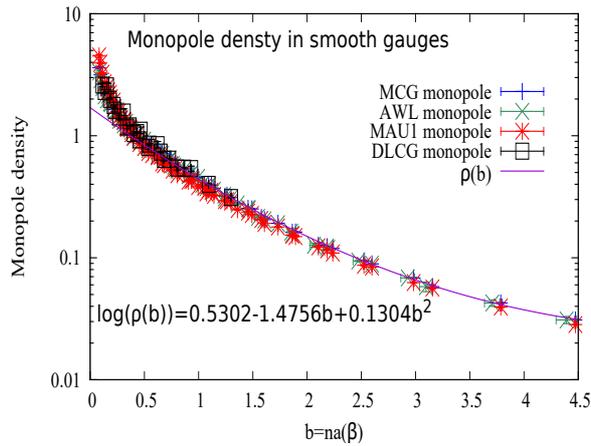}
    \caption{Comparison of the Abelian-like monopole densities versus $b=na(\beta)$ in MCG, AWL, DLCG and MAU1 cases. A uniform scaling curve is obtained for all gauges. }\label{density}
  \end{figure}

We evaluated a gauge-invariant density of the $n$-blocked monopole: 
\begin{eqnarray*}
\rho(a(\beta),n)=\frac{\sum_{\mu,s_n}\sqrt{\sum_a(k_{\mu}^{(n)a}(s_n))^2}}{4\sqrt{3}V_nb^3},
\end{eqnarray*}
which is a scale-invariant quantity depending on $a(\beta)$ and $n$ generally. But if we plot $\rho$ versus $b=na(\beta)$, we get a beautiful universal scaling function for all different gauge-fixings  
as shown in Fig.\ref{density}. Namely we obtained clear scaling behaviors $\rho(b)$ up to the 12-step blocking transformations for $\beta=3.0\sim 3.9$. Hence for fixed $b$, if we take $n\to \infty$, $a(\beta)\to 0$ that is we go to the continuum limit. The same beautiful scaling behaviors are obtained also in the case of the effective monopole action~\cite{Suzuki:2017zdh}. In addition to the scaling behaviors, the obtained scaling function is the same for four different gauges. Gauge independence is shown as naturally expected in the continuum limit.

\section{ACKNOWLEDGMENTS}
The authors would like to thank Y. Koma for giving them the computer  code of the multilevel method. The numerical simulations of this work were done using High Performance Computing resources at Cybermedia Center and Research Center for Nuclear Physics  of Osaka University,  at Cyberscience Center of Tohoku University and at KEK. The authors would like to thank these centers for their support of computer facilities. T.S  was finacially supported by JSPS KAKENHI Grant Number JP19K03848.

\end{document}

%% file: revmacro.tex
\definecolor{cyan}{cmyk}{1,0,0,0}
\definecolor{lightcyan}{cmyk}{0.5,0,0,0}
\definecolor{pastelcyan}{cmyk}{0.25,0,0,0}
\definecolor{magenta}{cmyk}{0,1,0,0}
\definecolor{yellow}{cmyk}{0,0,1,0}
\definecolor{lightyellow}{cmyk}{0,0,0.5,0}
\definecolor{pastelyellow}{cmyk}{0,0,0.25,0}
\definecolor{black}{cmyk}{0,0,0,1}
\definecolor{darkgray}{cmyk}{0,0,0,0.75}
\definecolor{gray}{cmyk}{0,0,0,0.5}
\definecolor{lightgray}{cmyk}{0,0,0,0.25}
\definecolor{white}{cmyk}{0,0,0,0}
\definecolor{red}{cmyk}{0,1,1,0}
\definecolor{orange}{cmyk}{0,0.5,1,0}
\definecolor{scarlet}{cmyk}{0,1,0.5,0}
\definecolor{brown}{cmyk}{0.5,0.75,1,0}
\definecolor{camel}{cmyk}{0.25,0.375,0.5,0}
\definecolor{cream}{cmyk}{0,0.2,0.3,0}
\definecolor{green}{cmyk}{1,0,1,0}
\definecolor{lightgreen}{cmyk}{0.5,0,0.5,0}
\definecolor{pastelgreen}{cmyk}{0.25,0,0.25,0}
\definecolor{mossgreen}{cmyk}{0.64,0.4,1,0}
\definecolor{yellowgreen}{cmyk}{0.5,0,1,0}
\definecolor{skyblue}{cmyk}{0.4,0.16,0,0}
\definecolor{royal}{cmyk}{1.0,0.5,0,0}
\definecolor{navyblue}{cmyk}{0.9,0.75,0.5,0}
\definecolor{blue}{cmyk}{1,1,0,0}
\definecolor{lightblue}{cmyk}{0.5,0.5,0,0}
\definecolor{lavender}{cmyk}{0.25,0.25,0,0}
\definecolor{violet}{cmyk}{0.75,1,0.25,0}
\definecolor{purple}{cmyk}{0.5,1,0.5,0}
\definecolor{pink}{cmyk}{0,0.5,0,0}
\definecolor{pastelpink}{cmyk}{0,0.25,0,0}
\def\gtsim{\mathrel{\hbox{\raise0.2ex
\hbox{$>$}\kern-0.75em\raise-0.9ex\hbox{$\sim$}}}}
\def\ltsim{\mathrel{\hbox{\raise0.2ex
\hbox{$<$}\kern-0.75em\raise-0.9ex\hbox{$\sim$}}}}
\def\llt{\mathrel{<\kern-0.5em <}}   
\def\ggt{\mathrel{>\kern-0.5em >}}   

\def\Tr{{\rm Tr}}
\def\del{{\partial}}

\def\kslash{k\kern-0.55em\raise 0.14ex\hbox{/}}
\def\Aslash{A\kern-0.6em\raise 0.14ex\hbox{/}}
\def\dslash{\del\kern-0.55em\raise 0.14ex\hbox{/}}
\def\therefore{ \hskip 0.0pt \raise0.1ex\hbox{.} \mkern -0.24mu \raise 1.1ex\hbox{.} \mkern -0.25mu \raise 0.1ex\hbox{.} \   } 
\def\because{ \hskip 0.0pt \raise 1.1ex\hbox{.} \mkern -0.24mu \raise 0.1ex\hbox{.} \mkern -0.25mu \raise 1.1ex\hbox{.} \  } 